\documentclass{jpsj2}
\title{
Griffiths Inequalities for Ising Spin Glasses on 
the Nishimori Line
}

\author{Hidetsugu \textsc{Kitatani}}

\inst{Department of General Education, Nagaoka University of
Technology, Nagaoka, Niigata 940-2188}

\abst{
The Griffiths inequalities for Ising spin glasses are proved on the Nishimori
line with various bond randomness which includes Gaussian and $\pm J$ bond randomness.
The proof for Ising systems with Gaussian bond randomness has already 
been carried out by Morita et al,
which uses not only the gauge theory but also
the properties of the Gaussian distribution, so that it cannot be directly
applied to the systems with other bond randomness. The present proof essentially
uses only the gauge theory, so that it does not depend on the detail properties of 
the probability distribution of random interactions. Thus,  
the results obtained from the inequalities for 
Ising systems with Gaussian bond randomness 
do also hold for those with various bond
randomness, especially with $\pm J$ bond randomness.
}

\kword{spin glass, Griffiths inequality, Ising model, Nishimori line, gauge theory,
thermodynamic limit}
\begin{document}
\maketitle

\section{Introduction}

The Griffiths inequalities make significant contributions to the 
understanding of phase transition for ferromagnetic Ising models.$^{1)}$
For one formulation, the  Griffiths inequalities are written as
\begin{equation}
\frac {d P}{d J_B} \ge 0,
\end{equation}
and
\begin{equation}
\frac {d \langle S_C \rangle}{d J_B} \geq 0,
\end{equation}
where $J_B$ is a positive interaction  and $S_C=\prod_{i \in C} S_i(S_i=\pm 1)$
is a product of Ising variables for arbitrary subset of the sites of the system.
The above inequalities state that the pressure, $P(=\log Z)$, and the  correlation
function, $\langle S_C \rangle $, are  monotonic increasing functions of
the strength of any interaction, $J_B$.
Using the two inequalities, various results can be proved, for example, the existence of the
thermodynamic limit of the pressure per  unit volume and correlation functions. 
They also give significant insights for the existence of
ferromagnetic phase transition.

For random Ising  systems which have both ferromagnetic and antiferromagnetic interactions, 
inequalities analogous to the first 
Griffiths inequality have been proved with various bond randomness.$^{2-4)}$
In  these cases, $J_B$ is a random variable, so that
it has been proved that the  pressure  is a monotonic increasing function
with respect to some parameter which controls the effect of an
interaction term.
Similar to the ferromagnetic cases, the inequalities may, for example, be used for
 the proof of the existence of the thermodynamic
limit of the pressure density.

As far as we know, however, the inequality analogous to the second
 Griffiths inequality has only been proved
on the Nishimori line, which is a restricted region in the phase diagram
 of random bond systems.
Morita et al  first proved the inequalities analogous  to both
the first and second Griffiths inequalities on the Nishimori line for
Ising systems with
Gaussian bond randomness.$^{5,6)}$
Using the two inequalities, the existence of the thermodynamic limit for
the pressure per unit volume and  correlation functions has been  proved
under various boundary conditions. Relations between the location of 
multicritical points for various lattices have also been derived.
The proof, however, uses not only the gauge theory, but also the properties 
of the Gaussian distribution, so that the proof
cannot be directly applied to the systems with other bond randomness,
for example, $\pm J$ bond
randomness.

Almost all the rigorous results obtained on the Nishimori line are proved, in 
essential,
only using the properties of the gauge transformation, so that they hold
not only for any lattice structure, dimension of the lattice, and range of
the interaction, but also for various bond randomness.$^{7,8)}$
For example, related to the second Griffiths inequality,
the present author  proved that the following inequality holds
 for finite $\pm J$ Ising
systems, $A$ and $B$, on the Nishimori line:
\begin{equation}
 [\langle S_0S_r \rangle]_{A} \geq [\langle S_0S_r \rangle]_{B},
\end{equation}
when the system $A$ is obtained from the system $B$ by adding random bonds.$^{9)}$
The above proof was originally  carried out for the $\pm J$ Ising models, and
used for deriving an inequality about the
 location of multicritical points of various lattices.
 However, the same argument can be applied
to the systems with other bond randomness, since the proof only uses the gauge theory.

Therefore, it is a natural question  whether 
the existence of the two Griffiths inequalities on the Nishimori line
is a special property of Ising systems with Gaussian bond randomness, or they hold
 without the detail property of the probability distribution of random interactions.
The main results of the present paper is that the two Griffiths inequalities do hold on
the Nishimori line for Ising systems with various bond randomness which includes
both Gaussian and $\pm J$ bond randomness, where
the proof essentially uses only the properties of the  
gauge transformation. Thus,  the results obtained from the two inequalities 
for Ising systems 
with Gaussian bond randomness$^{4,5)}$
do also hold for Ising systems with bond randomness considered in the present paper,
especially, for the systems with $\pm J$ bond randomness.

\section{The Model and the Bond Randomness}

Let us first define several quantities.
We treat the Ising spin system described by the Hamiltonian
\begin{equation}
 {\cal H} = -\sum_{A\subset V}J_{A}S_A,
  \label{Hamiltonian}
\end{equation}
where 
\begin{equation}
S_A=\prod_{i \in A}S_i.
\end{equation}
Here $V$ is the set of sites, and the sum over $A$ runs over all subsets
of $V$ among which interactions exist.
The number of sites in $A$ is arbitrary and may be different from
subset to subset.
The lattice structure is assumed to be reflected in the choice of $A$
for which $J_A\ne 0$.
The partition function of this system is written as
\begin{equation}
Z=\sum_{\{S_i\}}\exp (\sum_{A \subset \Omega}\beta_A J_A S_A),
\end{equation}
where we introduce  local inverse temperature, $\beta_A$. 
The reason is that we use local inverse temperature, $\beta_A$,  as a parameter which
controls the effect of  the interaction term, $-J_AS_A$, along the Nishimori line, as will be
shown later.
 To investigate the property of a physical quantity at inverse temperature, $\beta$,
 we must, of course, set all the local inverse temperature to $\beta_A=\beta$. The following argument,
however, does not depend on the value of  each local inverse temperature, $\beta_A$.

Then the pressure of the system is written as
\begin{equation}
P=[\log Z],
\end{equation}
where the  configurational average over the distribution of bond randomness 
is written as $[ \cdots ]$.

The probability distribution of the random interaction $J_A$ is denoted by $P(J_{A})$.
In this paper, we investigate the case that the probability distribution, $P(J_{A})$, 
satisfies the following two conditions:
\begin{equation}
P(-J_A)=P(J_A)\exp (-2\beta_{p,A}J_A)
\end{equation}
and
\begin{equation}
\frac {\partial P(J_A)}{\partial \beta_{p,A}}=(J_A-[J_A])P(J_A),
\end{equation}
where $\beta_{p,A}$ is a parameter which characterizes $P(J_A)$.

Most of the probability distributions of random interactions investigated 
in the spin glass problems may satisfy the above two conditions.
In the case of  $\pm J$ distribution  with the ferromagnetic bond 
concentration,
$p_A (p_A>1/2)$,  defining $\beta_{p,A}$ as
\begin{equation}
\exp (2\beta_{p,A}J)=\frac {p_A}{1-p_A},~~~~~(J>0)
\end{equation}
 we may rewrite $P(J_A)$ as
\begin{eqnarray}
P(J_A)&=&
p_A\delta(J_A-J)+(1-p_A)\delta(J_A+J)  \nonumber \\
&=&
\frac {\exp (\beta_{p,A} J)}{2\cosh (\beta_{p,A} J)}\delta (J_A-J)
+\frac {\exp (-\beta_{p,A} J)}{2\cosh (\beta_{p,A} J)}\delta (J_A+J) \\ \nonumber
&=&
\frac {1}{2\cosh (\beta_{p,A} J)}\left(\delta (J_A-J)
+\delta (J_A+J)\right)\exp(\beta_{p,A}J_A).
\end{eqnarray}
For  the Gaussian distribution with average, $J_{0,A}$, and variance, $\sigma_A$,
defining $\beta_{p,A}$ as
\begin{equation}
 \beta_{p,A}=\frac{J_{0,A}}{\sigma_A^2}
\end{equation}
we have
\begin{eqnarray}
P(J_A)&=&\frac {1}{\sqrt{2\pi} \sigma_A}\exp \left(\frac {-(J_A-J_{0,A})^2}
{2\sigma_A^2}\right)\\ \nonumber
&=&\frac {1}{\sqrt{2\pi} \sigma_A}\exp \left(\frac {-\sigma_A^2\beta_{p,A}^2}{2}\right)
\exp \left(\frac {-J_A^2}{2\sigma_A^2}\right)\exp (\beta_{p,A}J_A).
\end{eqnarray}
In both cases, we can easily see that $P(J_A)$ satisfies two
conditions, eqs. (8) and (9), by direct calculations.
It is noted that, in the case of  $\pm J$ distribution, $[J_A]=J\tanh(\beta_{p,A}J)$.

In the present notation, the interaction term, $-J_AS_A$, satisfies the Nishimori condition
when 
\begin{equation}
\beta_{p,A}=\beta_A.
\end{equation}
Namely, when eq. (14) is satisfied for all the subsets $\{A\}$, we may use  the properties
on the Nishimori line 
which have originally been proved by Nishimori, using the local gauge transformation.$^{7,8)}$

Here, we  show two properties of the probability distribution, $P(J_A)$,
which are often used in the rest of the paper.
When the function, $f(J_A)$, is an odd function of $J_A$, namely,
$f(-J_A)=-f(J_A)$, we obtain
\begin{equation}
[f(J_A)]=[\tanh(\beta_{p,A}J_A)f(J_A)].
\end{equation}
Also, when the function, $f(J_A)$, is an even function of $J_A$, and 
monotonic increasing function of  $\mid J_A \mid$, we have
\begin{equation}
[J_A\tanh(\beta_AJ_A)f(J_A)] \geq [J_A\tanh(\beta_AJ_A)][f(J_A)]
\end{equation}
The proofs of eq. (15) and ineq. (16) are shown in  Appendices A and B, respectively.

\section{ The inequalities on the Nishimori line}
Let us clarify the situation. We assume that all the interaction terms satisfy
the Nishimori condition, eq. (14).
In the following, we investigate the change of 
the pressure and the correlation functions with respect to arbitrary $\beta_B$, 
and prove the following two inequalities:
\begin{equation}
\frac {d}{d\beta_B}[P] \geq 0,
\end{equation}
and
\begin{equation}
\frac{d}{d\beta_B}[\langle S_C \rangle] \geq 0,
\end{equation}
where we denote the thermal average by angular brackets as $\langle \cdots \rangle$.
The above inequalities state that the
configurational average of the pressure and the
correlation functions are  monotonic increasing functions of $\beta_B$
on the Nishimori line. Here, we explain the role of $\beta_B$. When $\beta_B$
is zero,  there is no interaction term, $-J_BS_B$, in the system.
Increasing the value of $\beta_B$ makes the effect of the interaction term, $-J_BS_B$, larger,
though we consider the restricted case that $\beta_B$ always satisfies 
the Nishimori condition, $\beta_B=\beta_{p,B}$.
Namely, increasing  $\beta_B$ for the present system corresponds to 
increasing the strength of an interaction
for  ferromagnetic Ising models. Thus, on the Nishimori line,
ineqs. (17) and (18), 
play the same role as the
two Griffiths inequalities do for ferromagnetic Ising models.

Next, we briefly explain  the procedures of the proof, since the proof
of two inequalities can be carried out similarly.

\subsection{The basic transformation of the physical quantity}
First, using the identity,
\begin{equation}
\exp (\beta_B J_B S_B )
= \cosh (\beta_B J_B)(1+\tanh (\beta_B J_B) S_B),
\end{equation}
we rewrite the physical quantity, $Q$, so that the dependence of $\beta_B$ may be seen
explicitly. Corresponding to this procedure, we denote the configurational average over the distribution of bond randomness  except $J_B$
 as $[ \cdots ]^{'}$. The thermal average without the Boltzmann factor, $\exp(\beta_BJ_BS_B)$,
 is also denoted as $\langle \cdots \rangle^{'}$.

\subsection{The physical quantity on the Nishimori line}
Next, we rewrite the physical quantity, $[Q]$, on the Nishimori line,
 using the local gauge transformation.
Since we investigate the change of the physical quantity,
$[Q]$,  in the situation where
all the interaction terms  including $J_B$ always satisfy the Nishimori condition, 
we may use the properties on the Nishimori line, for example, 
$[\langle S_C \rangle ]=[
\langle S_C \rangle^{2}]$,
 in any step of the proof. Actually, we perform the local gauge transformation
  for all the spin variables, and
interactions except $J_B$. This can be done, since, in this step, there is no
Boltzmann factor, $\exp (\beta_B J_B S_B)$, and the dependence with respect to $J_B$ is
explicitly seen, as will be shown later.

\subsection{The change of the physical quantity on the Nishimori line}
Finally, we consider the change of the physical quantity, $[Q]$, with respect  to $\beta_B$.
Using eq. (9), we rewite the total derivative of $[Q]$ by $\beta_B$ as
\begin{eqnarray}
\frac{d}{d\beta_B}[Q]&=&\frac{d}{d\beta_B}
\int_{-\infty}^{\infty}dJ_BP(J_B)[Q]^{'}
\nonumber \\
&=&\int_{-\infty}^{\infty}dJ_B\left(\frac{\partial}{\partial \beta_B}
P(J_B)\right)[Q]^{'}+\int_{-\infty}^{\infty}dJ_BP(J_B)\left[\frac{\partial}{\partial \beta_B}Q\right]^{'}
\nonumber \\
&=&
\left[(J_B-[J_B])Q\right]+\left[\frac{\partial}{\partial \beta_B}Q\right].
\end{eqnarray}
We evaluate the second term of rhs  by direct calculations.
For the first term of rhs, when $Q$ is  an even function of $J_B$
and monotonic increasing function of $\mid J_B\mid$,
we have
\begin{equation}
\left[(J_B-[J_B])Q\right]=\left[(J_B\tanh(\beta_BJ_B)-[J_B\tanh(\beta_BJ_B)])Q\right]\geq 0,
\end{equation}
where we use  eq. (15) and ineq. (16). It must be noted that, for the $\pm J$ Ising models,
Q mentioned above becomes constant with respect  to $J_B$, since $J_B^{2}=J^2$, so that eq. (21) takes the value,
zero.

\section{The proof of the first inequality}
In this section, we prove the first inequality (17) following the procedures
shown in the previous section.

\subsection{The basic transformation of the pressure}
Using the identity (19),
we rewrite the pressure $[P]$ as
\begin{equation}
[P]=[\log Z]
=
[\log Z~{'}]^{'}+\left[\log (\cosh (\beta_B J_B))\right]+
\left[\log (1+\tanh (\beta_B J_B)\langle S_B \rangle^{'})\right].
\end{equation}
Here, $Z^{'}$ denotes the partition function without the Boltzmann factor, $\exp(\beta_B J_B S_B)$. 

\subsection{The pressure on the Nishimori line}
We expand the third term of rhs of eq.(22) as:
\begin{equation}
\left[\log(1+\tanh (\beta_B J_B)\langle S_B \rangle^{'})\right]
=\sum_{n=1}^{\infty}(-1)^{n-1}\frac{1}{n}\left[
\tanh^{n}(\beta_BJ_B)\langle S_B \rangle^{'n}\right].
\end{equation}
On the Nishimori line, using the local gauge transformation and eq. (15),
we can rewrite each odd term of the  series as
\begin{equation}
\left[
\tanh^{2n-1}(\beta_BJ_B)\langle S_B \rangle^{'(2n-1)}\right]
=
\left[
\tanh^{2n}(\beta_BJ_B)\langle S_B \rangle^{'2n}\right],
\end{equation}
where, the local gauge transformation is performed
for all the spin variables and all the interactions except $J_B$.
Then, we can further rewrite the third term of  rhs of  eq.(22) as
\begin{eqnarray}
\left[\log(1+\tanh (\beta_B J_B)\langle S_B \rangle^{'})\right]
&=&\left[\sum_{n=1}^{\infty}\frac{1}{2n(2n-1)}\tanh^{2n}(\beta_BJ_B)\langle S_B \rangle^{'2n}
\right] 
\nonumber \\
&=&[F(J_B)],
\end{eqnarray}
were we define $F(J_B)$ as
\begin{eqnarray}
F(J_B)&=&\frac {1}{2}\left(
\left(1+\tanh (\beta_B J_B)\langle S_B \rangle^{'}\right)
\log \left(1+\tanh (\beta_B J_B)\langle S_B \rangle^{'}
\right)
\right) \nonumber \\
&+&
\frac {1}{2}
\left(\left(1-\tanh (\beta_B J_B)\langle S_B \rangle^{'}
\right)\log \left(1-\tanh (\beta_B J_B)\langle S_B \rangle^{'}
\right)
\right).
\end{eqnarray}
By direct calculation, we have
\begin{equation}
F(-J_B)=F(J_B).
\end{equation}
We also obtain
\begin{equation}
\frac {\partial}{\partial J_B}F(J_B)=\frac{\beta_B \langle S_B\rangle^{'}}{2\cosh^{2}(\beta_BJ_B)}
\log \left(\frac{1+\tanh (\beta_B J_B)\langle S_B\rangle^{'}}
{1-\tanh (\beta_B J_B)\langle S_B\rangle^{'}}\right)\geq 0, ~~~~~(J_B>0)
\end{equation}
regardless of the sign of $\langle S_B\rangle^{'}$.
Namely, $F(J_B)$ is an even function of $J_B$ and monotonic increasing function
of $\mid J_B \mid$.

\subsection{The change of the pressure on the Nishimori line}
Now, we calculate the total derivative of the pressure, $[P]$, by
$\beta_B$ on the Nishimori line.
First, we have
\begin{equation}
\frac {d}{d\beta_B}[P]=\frac {d}{d\beta_B}[\log (\cosh (\beta_B J_B))]
+\frac {d}{d\beta_B}\left[\log (1+\tanh (\beta_B J_B)\langle S_B \rangle^{'})
\right].
\end{equation}
The first term of rhs of eq. (29) is directly calculated as
\begin{eqnarray}
\frac {d}{d\beta_B}[\log (\cosh (\beta_B J_B))]
&=&[(J_B-[J_B])\log (\cosh (\beta_B J_B))]+[J_B\tanh (\beta_B J_B)] \nonumber\\
&\geq&0,
\end{eqnarray}
where we use ineq. (21), since $\log (\cosh (\beta_B J_B))$
is an even function of $J_B$, and monotonic increasing function of $\mid J_B \mid$.

Similarly, the second term of rhs of eq. (29) is calculated as
\begin{equation}
\frac {d}{d\beta_B}\left[\log(1+\tanh (\beta_B J_B)\langle S_B \rangle^{'})
\right]
=\left[(J_B-[J_B])
F(J_B)\right]+\left[ \frac {\partial}{\partial \beta_B}F(J_B)\right].
\end{equation}
Using, ineq. (21), it yields that
\begin{equation}
[(J_B-[J_B)])F(J_B)
] \ge 0
\end{equation}
For the second term of rhs of eq. (31), a direct calculation gives
\begin{equation}
\left[ \frac {\partial}{\partial \beta_B}F(J_B)\right]=
\left[\frac{J_B \langle S_B\rangle^{'}}{2\cosh^{2}(\beta_BJ_B)}
\log \left(\frac{1+\tanh (\beta_B J_B)\langle S_B\rangle^{'}}
{1-\tanh (\beta_B J_B)\langle S_B\rangle^{'}}\right)\right]
\geq0,
\end{equation}
since  it is easily seen that each term in the square brackets  of rhs of eq. (33)
is nonnegative regardless of the sign of $J_B\langle S_B \rangle^{'}$.
Thus, we obtain the first inequality (17).

\section{The proof of the second inequality}
We can similarly prove the second inequality (18), which states that the correlation function
is a monotonic increasing function of $\beta_B$.

\subsection{The basic transformation of the correlation function}
Using the identity (19), we  rewrite the correlation function as
\begin{equation}
\langle S_C\rangle 
=\frac {\langle S_C \rangle^{'}
+\tanh (\beta_B J_B)\langle S_B S_C \rangle^{'}}
{1+\tanh (\beta_B J_B)\langle S_B \rangle^{'}}.
\end{equation}
In this subsection, for simplicity, we define $x_B$ and $x_{p.B}$ as
\begin{equation}
x_B=\tanh (\beta_B J_B),
\end{equation}
and
\begin{equation}
x_{p,B}=\tanh (\beta_{p,B}J_B).
\end{equation}
We further rewrite the correlation function as
\begin{eqnarray}
\langle S_C \rangle
&=& \frac {\langle S_C \rangle^{'}
+x_B(\langle S_BS_C\rangle^{'}
-\langle S_B\rangle^{'}\langle S_C\rangle^{'})
-x_B^2\langle S_BS_C\rangle^{'}
\langle S_B \rangle^{'}}
{1-x_B^2\langle S_B \rangle^{'2}} \nonumber \\
&=&
\langle S_C \rangle^{'}
+\frac {x_B(\langle S_BS_C\rangle^{'}
-\langle S_B\rangle^{'}\langle S_C\rangle^{'})
-x_B^2(\langle S_BS_C\rangle^{'}
-\langle S_B\rangle^{'}\langle S_C\rangle^{'})
\langle S_B \rangle^{'}
}
{1-x_B^2\langle S_B \rangle^{'2}}.
\end{eqnarray}
Then, the configurational average of the correlation function
can be written as
\begin{eqnarray}
[\langle S_C \rangle]
&=&[\langle S_C \rangle^{'}]^{'}
+\left[
\frac{(\langle S_BS_C\rangle^{'}
-\langle S_B\rangle^{'}\langle S_C\rangle^{'})
(x_B-x_B^2\langle S_B \rangle^{'})}
{1-x_B^2\langle S_B \rangle^{'2}}
\right] \nonumber \\
&=&
[\langle S_C \rangle^{'}]^{'}
+\left[
\frac{(\langle S_BS_C\rangle^{'}
-\langle S_B\rangle^{'}\langle S_C\rangle^{'})
(x_{p,B}x_B-x_B^2\langle S_B \rangle^{'})}
{1-x_B^2\langle S_B \rangle^{'2}}
\right],
\end{eqnarray}
where we use eq. (15) for the second term of rhs.

\subsection{The correlation function on  the Nishimori line}
On the Nishimori line, since $\beta_{p,B}=\beta_B$, and $x_{p,B}=x_B$,
the correlation function can
be rewritten as
\begin{equation}
[\langle S_C \rangle]
=[\langle S_C \rangle^{'}]^{'}
+\left[
\frac{x_B^2(\langle S_BS_C\rangle^{'}
-\langle S_B\rangle^{'}\langle S_C\rangle^{'})(1-\langle S_B \rangle^{'})}
{1-x_B^2\langle S_B \rangle^{'2}}
\right].
\end{equation}
Performing  the local gauge transformation for all the spin variables and
interactions except $J_B$, we obtain
\begin{equation}
\left[
\frac{x_B^2\langle S_BS_C\rangle^{'}
}
{1-x_B^2\langle S_B \rangle^{'2}}
\right]=
\left[
\frac{x_B^2\langle S_BS_C\rangle^{'2}
}
{1-x_B^2\langle S_B \rangle^{'2}}
\right],
\end{equation}
\begin{equation}
\left[
\frac{x_B^2
\langle S_B\rangle^{'}\langle S_C\rangle^{'}}
{1-x_B^2\langle S_B \rangle^{'2}}
\right]
=
\left[
\frac{x_B^2\langle S_BS_C\rangle^{'}
\langle S_B\rangle^{'}}
{1-x_B^2\langle S_B \rangle^{'2}}
\right]
=
\left[
\frac{x_B^2\langle S_BS_C\rangle^{'}
\langle S_B\rangle^{'}\langle S_C\rangle^{'}}
{1-x_B^2\langle S_B \rangle^{'2}}
\right],
\end{equation}
and
\begin{equation}
\left[
\frac{x_B^2
\langle S_B\rangle^{'2}\langle S_C\rangle^{'}}
{1-x_B^2\langle S_B \rangle^{'2}}
\right]
=
\left[
\frac{x_B^2
\langle S_B\rangle^{'2}\langle S_C\rangle^{'2}}
{1-x_B^2\langle S_B \rangle^{'2}}
\right].
\end{equation}
Thus, the second term of rhs of eq. (39) becomes
\begin{equation}
\left[
\frac{x_B^2(\langle S_BS_C\rangle^{'}
-\langle S_B\rangle^{'}\langle S_C\rangle^{'})(1-\langle S_B \rangle^{'})}
{1-x_B^2\langle S_B \rangle^{'2}}
\right]
=\left[ \frac{x_B^2(\langle S_BS_C\rangle^{'}
-\langle S_B\rangle^{'}\langle S_C\rangle^{'})^2}
{1-x_B^2\langle S_B \rangle^{'2}}
\right].
\end{equation}
Hence, we obtain
\begin{equation}
[\langle S_C \rangle]
=[\langle S_C \rangle^{'}]^{'}
+\left[ \frac{\tanh^{2}(\beta_BJ_B)(\langle S_BS_C\rangle^{'}
-\langle S_B\rangle^{'}\langle S_C\rangle^{'})^2}
{1-\tanh^{2}(\beta_BJ_B)\langle S_B \rangle^{'2}}
\right],
\end{equation}
where, for the second term of rhs of eq. (44),
it is easily seen that the term in the square brackets is 
an even function of $J_B$
and monotonic increasing function of $\mid J_B\mid$.

\subsection{The change of the correlation function on the Nishimori line}
The total derivative of the correlation function by $\beta_B$ 
on the Nishimori line can be written as
\begin{eqnarray}
\frac{d}{d\beta_B}[\langle S_C \rangle]
&=&\left[(J_B-[J_B])\frac{\tanh^{2}(\beta_BJ_B)(\langle S_BS_C\rangle^{'}
-\langle S_B\rangle^{'}\langle S_C\rangle^{'})^2}
{1-\tanh^{2}(\beta_BJ_B)\langle S_B \rangle^{'2}}
\right] \nonumber \\
&+&\left[\frac {\partial}{\partial \beta_B}\left( \frac{\tanh^{2}(\beta_BJ_B)(\langle S_BS_C\rangle^{'}
-\langle S_B\rangle^{'}\langle S_C\rangle^{'})^2}
{1-\tanh^{2}(\beta_BJ_B)\langle S_B \rangle^{'2}}
\right)
\right]
\end{eqnarray}
Using ineq. (21), the first term of rhs of eq. (45) becomes nonnegative.
For the second term of rhs of eq. (45), a direct calculation gives
\begin{eqnarray}
&&\left[\frac {\partial}{\partial \beta_B}\left( \frac{\tanh^{2}(\beta_BJ_B)(\langle S_BS_C\rangle^{'}
-\langle S_B\rangle^{'}\langle S_C\rangle^{'})^2}
{1-\tanh^{2}(\beta_BJ_B)\langle S_B \rangle^{'2}}
\right)
\right] \\
&=&
\left[ \frac{2J_B\tanh(\beta_BJ_B)(\langle S_BS_C\rangle^{'}
-\langle S_B\rangle^{'}\langle S_C\rangle^{'})^2}
{\cosh^{2}(\beta_BJ_B)(1-\tanh^{2}(\beta_BJ_B)\langle S_B \rangle^{'2})^2}
\right] \geq 0
\end{eqnarray}
Thus, we obtain the second inequality (18).

\section{Summary and discussion}
We have proved two inequalities, ineqs. (17) and (18), for
 Ising spin glasses on the Nishimori line with various bond randomness which
includes Gaussian and $\pm J$ bond randomness,
where the probability distribution of random interactions 
must satisfy two conditions, eqs. (8) and (9).  The two inequalities, which correspond to
the Griffiths inequalities for ferromagnetic Ising models, state
that, along the Nishimori line,  the pressure and the correlation functions are monotonic increasing functions
of any $\beta_B$ which controls the effect of the interaction term, $-J_BS_B$.
The present results are an generalization of those by
Morita et al for Ising systems with Gaussian bond randomness,$^{5)}$ where the  proof 
uses not only the gauge theory but also the properties of the
Gaussian distribution, so that it cannot be directly applied to the
systems with other bond randomness. The present proof  essentially
uses only the gauge theory, so that it holds without the detail property of the 
probability distribution
of random interactions.

Using the present proof, the results obtained from the
two inequalities for Ising system with Gaussian
bond randomness$^{5,6)}$ can be also derived for the systems with various bond randomness.
In the research of  the Ising spin glass problems, however, most studies have been 
carried out
for the systems with Gaussian or $\pm J$ bond randomness.
Thus, we may insist that the most  important physical consequence of the
present paper is that it is found that the results obtained for Ising systems with Gaussian bond
randomness$^{5,6)}$ do also hold for Ising systems with $\pm J$ bond randomness.

Let us briefly explain several physical consequences on the
Nishimori line for  regular lattices which can be derived 
 from the two inequalities, though
they have   already been explained by Morita et al for Ising systems with Gaussian bond
randomness.$^{5)}$

From the first inequality (17), we can show that the pressure 
has the well known super-additivity,$^{2,3)}$ namely
\begin{equation}
[P]_{V} \ge \sum_i[ P]_{V_i},
\end{equation}
where $[P]_V$ denotes the pressure of the set of sites, $V$, and
\begin{equation}
V =\sum_i V_i,
\end{equation}
since $[P]_V$ is obtained from $\sum_i [P]_{V_i}$ by adding random bonds among
 $V_i$ and $V_j$.
 Thus, 
we can show the existence of the thermodynamic limit of the pressure density
on the Nishimori line
under free boundary conditions, assuming invariance
by translation with respect to random interactions  and  the stability boundedness.$^{2.3)}$
For  $\pm J$ Ising spin glasses with short range interactions, it is
easily seen that the pressure has a definite stable boundedness.

From the second inequality (18), we can show the existence of the thermodynamic limit
of the correlation functions on the Nishimori line 
under free and fixed boundary conditions.
For free boundary conditions, when we consider two finite sets of sites, $V$ and $V^{'}$
 so that $V^{'} \supset V$, we get
\begin{equation}
[\langle S_C \rangle]_{V^{'}} \geq [\langle S_C \rangle]_{V},
\end{equation}
since  $V^{'}$ is obtained from  $V$ by adding  random bonds.
Thus, we may say that the correlation function is a monotonic increasing
function with
the system size. With the fact that the correlation function is bounded by unity,
we can assert the existence of the thermodynamic limit of the correlation function
under free boundary conditions.
We can also prove the existence of the thermodynamic limit of the correlation functions
under fixed boundary conditions by similar procedure. (See Ref. 5 for details.)

From the second inequality, we can also obtain  the relation between
the location of the multicritical points.
When the lattice $L_1$ is obtained from the lattice $L_2$ by adding random bonds,
we get the following inequality for the magnetization
\begin{equation}
[\langle S_i \rangle ]_{L_1} \geq [\langle S_i \rangle ]_{L_2},
\end{equation}
from which, we have
\begin{equation}
T_{{\rm c}L_1}\ge T_{{\rm c},L_2},
\end{equation} 
where $T_{{\rm c},L_1}(T_{{\rm c},L_2})$ is the temperature of the multicritical point of 
the lattice, $L_1(L_2)$.

Finally, we mention about the relation between  ineq. (3)  and the
second ineq. (18) of the present paper including the work by
Morita et al.$^{5)}$ Inequality (3) only states that, for finite systems, the correlation function of some
positive $\beta_B$ is larger than that of $\beta_B =0$, and cannot give the
information between  the values of   the correlation functions of two different positive
$\beta_B$. Compared to the above fact, from eq. (44),  we can  explicitly see
how the value of the correlation
funcion increases as  $\beta_B$ increases. 
Thus, ineq. (3) has clearly less information than ineq. (18), which comes from the derivative of
eq. (44) by $\beta_B$.
However, ineq. (50), for example, can be derived using only ineq. (3),
 though it was not explicitly  mentioned in Ref. 9. In the proof of ineq. (3),
 the condition for the probability distribution of random interactions is only one condition, namely,
 eq. (8), which states that the system has the Nishimori line itself.
 Thus, following the argument executed in Ref. 5,
the existence of the thermodynamic limit of the correlation function on the Nishimori line
 under free boundary conditions
 may be proved for all the systems which have the Nishimori line in the phase diagram.
 The relation  between the location of multicritical points of various lattices can also
 be derived for the same systems.

\section*{Acknowledgement}
We thank Satoshi Morita for useful comments.

\appendix
\section{Derivation of eq. (15)}

In this appendix we explain the derivation of eq. (15).
Through  Appendices A snd B, we denote the configurational average over the distribution of bond randomness  except $J_A$  as $[ \cdots ]^{'}$. 

 Using eq. (8) and the fact that $f(J_A)$ is an odd function of $J_A$,
we have
\begin{eqnarray}
[\exp(-\beta_{p,A}J_A)f(J_A)]
&=&\int^{\infty}_{-\infty}dJ_A P(J_A)\exp(-\beta_{p,A}J_A)[f(J_A)]^{'} \nonumber \\
&=&\int^{\infty}_{-\infty}dJ_A P(-J_A)\exp(\beta_{p,A}J_A)[f(J_A)]^{'}  \nonumber \\
&=&-\int^{-\infty}_{\infty}dJ_A P(J_A)\exp(-\beta_{p,A}J_A)[f(-J_A)]^{'} \nonumber \\
&=&-\int^{\infty}_{-\infty}dJ_A P(J_A)\exp(-\beta_{p,A}J_A)[f(J_A)]^{'} \nonumber \\
&=&-[\exp(-\beta_{p,A}J_A)f(J_A)],
\end{eqnarray}
which implies
\begin{equation}
[\exp(-\beta_{p,A}J_A)f(J_A)]=0.
\end{equation}
Thus, we also obtain
\begin{equation}
\left[\frac{\exp(-\beta_{p,A}J_A)}
{\exp(\beta_{p,A}J_A)+\exp(-\beta_{p,A}J_A)}f(J_A)\right]=0,
\end{equation}
since $(1/
(\exp(\beta_{p,A}J_A)+\exp(-\beta_{p,A}J_A))f(J_A))$ is also an odd function of $J_A$.
Therefore, it yields that
\begin{eqnarray}
[f(J_A)]&=&
[f(J_A)]-2\left[\frac{\exp(-\beta_{p,A}J_A)}
{\exp(\beta_{p,A}J_A)+\exp(-\beta_{p,A}J_A)}f(J_A)\right] \nonumber \\
&=&[\tanh(\beta_{p,A}J_A)f(J_A)]
\end{eqnarray}

\section{Derivation of ineq. (16)}
When two functions, $f(J_A)$ and $J_A\tanh(\beta_AJ_A)$, are both 
even functions of $J_A$ and monotonic increasing functions of 
$\mid J_A \mid$, we have
\begin{equation}
[(J_A\tanh(\beta_AJ_A)-J_0\tanh(\beta_AJ_0))(f(J_A)-f(J_0))] \geq 0,
\end{equation}
 for any constant value, $J_0$, from which, it yields that
\begin{equation}
[J_A\tanh(\beta_AJ_A)f(J_A)]\geq J_0\tanh(\beta_AJ_0)[f(J_A)]+\left([J_A\tanh(\beta_AJ_A)]-J_0\tanh(\beta_AJ_0)\right)[f(J_0)]^{'}.
\end{equation}
Here, we can choose the value, $J_0$,  so that it satisfies
\begin{equation}
[J_A\tanh(\beta_AJ_A)]=J_0\tanh(\beta_AJ_0).
\end{equation}
Substituting eq. (B.3) into eq.(B.2), we obtain ineq. (16).


\end{document}